**Title: Miniature Magnetic Nanoislands in a Morphotropic Cobaltite Matrix**

**Authors**: Shengru Chen,[1,2,†] Dongke Rong,[1,†] Yue Xu,[1,2] Miming Cai,[3] Xinyan Li,[1] Qinghua Zhang,[1] Shuai Xu,[1,2] Yan-Xing Shang,[1] Haitao Hong,[1,2] Ting Cui,[1,2] Qiao Jin,[1] Jia-Ou Wang,[4] Haizhong Guo,[5] Lin Gu,[6] Qiang Zheng,[7] Can Wang,[1,2,8] Jinxing Zhang,[3] Gang-Qin Liu,[1,2,8] Kuijuan Jin,[1,2,8,*] and Er-Jia Guo[1,2,*]

**Affiliations:**

[1] Beijing National Laboratory for Condensed Matter Physics and Institute of Physics, Chinese Academy of Sciences, Beijing 100190, China

[2] Department of Physics & Center of Materials Science and Optoelectronics Engineering, University of Chinese Academy of Sciences, Beijing 100049, China

[3] Department of Physics, Beijing Normal University, Beijing 100875, China

Key Laboratory of Material Physics & School of Physics and Microelectronics, Zhengzhou University, Zhengzhou 450001, China

[4] Institute of High Energy Physics, Chinese Academy of Sciences, Beijing 100049, China

[5] School of Physics and Microelectronics, Zhengzhou University, Zhengzhou 450001, China

[6] National Center for Electron Microscopy in Beijing and School of Materials Science and Engineering, Tsinghua University, Beijing 100084, China

[7] CAS Key Laboratory of Standardization and Measurement for Nanotechnology, CAS Center for Excellence in Nanoscience, National Centre for Nanoscience and Technology, Beijing 100190, China

[8] Songshan Lake Materials Laboratory, Dongguan, Guangdong 523808, China

E-mail: kjjin@iphy.ac.cn and ejguo@iphy.ac.cn

†These authors contribute equally to the manuscript.

**Abstract**

High-density magnetic memories are key components in spintronics, quantum computing, and energy-efficient electronics. Reduced dimensionality and magnetic domain stability at the nanoscale are essential for the miniaturization of magnetic storage units. Yet, inducing magnetic order, and selectively tuning spin–orbital coupling at specific locations have remained challenging. Here we demonstrate the construction of switchable magnetic nanoislands in a nonmagnetic matrix based on cobaltite homostructures. The magnetic and electronic states are laterally modified by epitaxial strain, which is regionally controlled by freestanding membranes. Atomically sharp grain boundaries isolate the crosstalk between magnetically distinct regions. The minimal size of magnetic nanoislands reaches ~35 nm in diameter, enabling an areal density of ~400 Gbit/in$^2$. Besides providing an ideal platform for precisely controlled read/write schemes, this methodology can enable scalable and patterned memories on silicon and flexible substrates for various applications.

**One sentence summary**: Switchable magnetic nanoislands achieved in a morphotropic cobaltite matrix can pave the way for scalable memory devices.

**Introduction**

Magnetic storage units are the smallest patterns of magnetization in magnetizable materials used for data storage. The digital information in the binary form (0 or 1 bit) can be accessed or processed from a magnetic head [1]. To increase the capacity of magnetic storage, the dimensionality of basic elements must be reduced while preserving the stability of magnetic nanodomains [2,3]. However, superparamagnetic effects in magnetic materials limit the minimum domain size [4,5]. Patterning continuous magnetic media composed of highly coupled magnetic grains into discrete single nanostructures has been proposed. Using direct-write e-beam lithography, magnetic circular dot arrays with an areal density of ~ 65 Gbit/in$^2$ have been commercialized [6]. However, they have reached not only the device fabrication limit but also the characterization limit of individual elements. Therefore, advanced materials and evolutionary fabrication processes are required to further reduce the periodicity of magnetic units to <25 nm to produce patterned media with enhanced areal density (>1 Tbit/in$^2$) [7].

Magnetic oxides are ideal candidates for fabricating memory devices owing to their chemical stability, strong magnetic anisotropy, and relatively large coercivity. Heavy orbital hybridization between transition metal ions and oxygen ions results in strong correlations among different degrees of freedom [8]. Previous studies demonstrated the manipulation of magnetic anisotropy owing to oxygen octahedral interconnection at interfaces [9-11]. For instance, the octahedral rotation in manganite layers is steeply suppressed by inserting a single-unit-cell-thick SrTiO$_3$ layer between films and substrates, resulting in the rotation of the in-plane magnetic easy axis [9,10]. Alternatively, the control of structural parameters can be achieved by capping an ultrathin layer with dissimilar symmetry [11]. An enhanced magnetic phase transition temperature and perpendicular magnetic anisotropy in ultrathin ferromagnetic oxides were observed after interfacial modifications [12]. Furthermore, using lithographic patterning, we locally controlled the magnetic properties of oxide heterostructures [13], expanding possibilities for developing memory devices at the nanoscale. Despite the control of magnetic anisotropy by oxide interface engineering, the crosstalk between nearby magnetic domains in thin films remains unavoidable. One strategy to break centrosymmetry in continuous two-dimensional (2D) thin films is to fabricate magnetizable domains in a nonmagnetic matrix using single materials. It requires magnetic materials that are highly sensitive to structural distortions so that their magnetic and electronic properties can be directly tuned.

Recent investigations revealed that the magnetic states of ferroelastic LaCoO$_3$ (LCO) are actively correlated to epitaxial strain [14]. Small lattice distortions such as deformations and oxygen octahedral tilts (or rotations) modify the balance between crystal field splitting ($\Delta_{cf}$) and intraatomic exchange interaction ($\Delta_{ex}$); thus, the spin states of Co ions are reversibly switched [15]. Generally, tensile strain increases the population of higher-spin-state Co$^{3+}$ ions, resulting in a robust ferromagnetic state, whereas in-plane compression promotes lower-spin-state transition; as a result, long-range magnetic ordering cannot be formed in compressively strained LCO films [16]. Therefore, LCO has been recognized as a promising route for controlling magnetic states using strain engineering. Typically, for modifying the strain states of epitaxial films, changing substrate materials or introducing buffer layers with different compositions was reported [17, 18]. However, this regulates the in-plane strain states, crystalline orientations, or rotation patterns of the films once a substrate is chosen. It is challenging to modify electronic and magnetic states regionally along the film plane. Recently, Wu *et al.* and Chen *et al.* reported the growth of ferroic oxide thin films with laterally tunable strains and orientations using acid-soluble manganite layers and water-soluble sacrificial layers, respectively [19,20]. In both cases, the original substrates, and suspended freestanding membranes served as independent building blocks for the epitaxial growth of functional oxides. Here, we report the construction of ferromagnetic nanoislands with diameters of tens of nanometers in nonmagnetic media based on cobaltite homostructures. The minimum domain size achieved using this approach enables the development of ultrahigh-density magnetic storage for various applications.

**Results**

We firstly deposited water-soluble sacrificial Sr$_3$Al$_2$O$_6$ (SAO) layers (≈30 nm) [21] and SrTiO$_3$ (STO) layers (≈3 nm) subsequently on (001)-oriented (LaAlO$_3$)$_{0.3}$–(Sr$_2$AlTaO$_6$)$_{0.7}$ (LSAT) single-crystalline substrates using pulsed laser deposition (PLD). The ultrathin STO membrane was thermally transferred on LaAlO$_3$ (LAO) and other substrates (glass, α-Al$_2$O$_3$, and silicon) after delamination from LSAT substrates (see Methods). The freestanding STO (FS-STO) membranes maintained their original shape and high crystallinity at the millimeter scale. To test the distinct physical properties, we covered the substrates partially with FS-STO membranes with an area ratio of 40%. Then, LCO films were fabricated on the modified substrates. Figure 1a shows the schematic of a grain boundary region in LCO hybrid homostructures. The LCO films (defined as $LCO_C$) grown directly on LAO substrates showed a compressive strain of ~ –0.52%, resulting in ~0.7% elongation along the out-of-plane direction (Figure 1b). On the other hand, the LCO films (defined as $LCO_T$) were tensile-strained by FS-STO membranes. Using both X-ray diffraction (XRD) θ–2θ scans and reciprocal space mappings (RSM), we obtained the lattice parameters of $LCO_T$ films grown directly on single-crystalline STO substrates and freestanding membranes (Figures 2b and 2c). We found that the in-plane lattice constant of FS-STO reduced to 3.874 ± 0.015 Å, while its out-of-plane lattice constant increased to 3.920 ± 0.008 Å. This result indicates that ultrathin FS-STO membranes are more responsive to in-plane compression than the relatively thick LCO top layers. We determined that the in-plane tensile strain in $LCO_T$ films reduced from 2.5% (on STO substrates) to 1.68% (on FS-STO membranes).

Microscopic structural analysis was examined by cross-sectional high-angle annular dark field (HAADF) imaging via scanning transmission electron microscopy (STEM) (Figure 1d and Figure S1). A representative HAADF-STEM image from a grain boundary region in LCO hybrid homostructures shows that the $LCO_C$/LAO and $LCO_T$/FS-STO heterointerfaces are atomically sharp. Both $LCO_C$ and $LCO_T$ films were epitaxially grown without apparent defects and chemical disorders. We calculated the atomic distance between A-site ions along in-plane (Figure 1e) and out-of-plane (Figure 1f) directions. $LCO_C$ ($LCO_T$) layers were coherently grown on LAO substrates (FS-STO membranes) with nearly identical in-plane lattice constants. The out-of-plane lattice constant of $LCO_T$ was considerably smaller than that of $LCO_C$, consistent with XRD results. The FS-STO membranes with a thickness of ~7 unit cells uniformly covered LAO substrates with an extremely narrow gap (~1 nm). Different from epitaxial thin films, the chemically unbonded FS-STO membranes did not involve misfit strain resulting from LAO substrates, allowing the suspended LCO films to follow the strain and orientation of FS-STO membranes (Figure S2). We observed mainly periodic dark stripe patterns perpendicular to the interface in $LCO_T$ layers while these superstructures were absent in $LCO_C$ layers. These results are in excellent agreement with previous observations [22]. Since the degree of tensile strain reduces in LCO$_T$ layers grown on FS-STO membranes, relatively fewer stripes appear compared to the LCO$_T$ films sandwiched between STO layers [23]. Notably, the dark stripes exist only in the regions close to FS-STO membranes, indicating that the tensile strain partially relaxes with increasing LCO$_T$ film thickness. Furthermore, we performed optical second-harmonic generation (SHG) measurements on LCO$_C$ and LCO$_T$ layers independently. As indicated in Figure 1g and Figure S3, LCO$_T$ layers fit the point group *P*4*mm*, while LCO$_C$ layers belong to the point group *m*, suggesting that LCO layers exhibit different crystallographic symmetries under distinct misfit strains.

The observed structural modification is strongly related to macroscopic magnetic properties in LCO hybrid homostructures. We performed magnetic measurements on LCO hybrid homostructures, LCO single-crystalline films grown on LAO and STO substrates. As shown in Figures 2a and 2b, the magnetization (*M*) exhibits a square-like hysteresis loop as a function of field and a sharp transition with increasing temperature, corroborating the ferromagnetic order in LCO hybrid homostructures. The *M*–*H* curves of LCO hybrid homostructures exhibit a small kink in the low-field region. This anomaly is attributed to the paramagnetic (PM) signal from LCO single-crystalline LCO films on LAO [24]. Considering the area ratio (LCO$_C$:LCO$_T$ = 60:40), the saturation *M* and field dependency of LCO hybrid homostructures are almost identical to those of LCO single films grown on STO (Figure S4). The distinct magnetic properties of LCO$_C$ and LCO$_T$ layers on a single LCO hybrid homostructure were revealed by performing nanodiamond (ND) nitrogen-vacancy (NV) magnetometry measurements [25]. NV magnetometry was performed at zero magnetic fields and switchable temperatures ranging from 6 K to 120 K. By recording optical detected magnetic resonance (ODMR) spectra of the numbers of NV centers in a single ND, we calculated the projection of the magnetic stray field along the NV axis due to the energy splitting (2γ*B*) in the presence of weak magnetic perturbation (*B*). Since NDs were dispersed on the sample surface, we measured the ODMR spectrum from either LCO$_T$ or LCO$_C$ layers at different locations independently. Figure 2c shows the ODMR spectra of a single ND on LCO$_T$ (left panels) and LCO$_C$ (right panels) at various temperatures. At low temperatures, the splitting between two resonance peaks from LCO$_T$ layers was large, and it reduced gradually with increasing temperature. However, the energy splitting from LCO$_C$ layers maintained a small but fixed value as the temperature varied from 6 to 120 K. We calculated 2γ$B$ by subtracting two resonant peak frequencies and summarized these values as open symbols in Figure 2b, although NV magnetometry cannot obtain the exact net moment of individual layers because the magnitude of 2γ$B$ depends on the number of NVs in a single ND, proximity distance, and magnetic homogeneity. The temperature-dependent 2γ$B$ yields the estimated Curie temperature ($T_C$) of LCO$_T$ layers (~80 K), which agreed with that obtained from SQUID measurements. 2γ$B$ obtained from LCO$_C$ layers remained nearly constant at all temperatures, suggesting that $M$ weakly depended on temperature when LCO films were compressively strained. The discrepancies between 2γ$B$–$T$ and $M$–$T$ curves at low temperatures indicate additional PM contributions from substrates and FS-STO membranes. The SQUID measures the total $M$ from an entire sample, while NV magnetometry is insensitive to PM signals originating from the layers/substrates under LCO$_C$ and LCO$_T$ layers, making it advantageous to apply NV magnetometry to measure magnetic properties in different regions.

Field-dependent magnetic force microscopy (MFM) measurements were conducted at 6 K. Figures 2d and 2e show the MFM images of LCO$_T$ and LCO$_C$ layers collected when magnetic fields switched between –2 T and 2 T, respectively. At –2 T, the LCO$_T$ layers contained static multiple magnetic domains with reversed polarity. As the magnetic field increased from –2 T to 2 T, magnetic domains (yellow) in LCO$_T$ shrank and showed reversed polarity (blue). These results demonstrate the ferromagnetic character of LCO$_T$ layers. On the contrary, LCO$_C$ layers did not show magnetic phase contrast, and these domains barely changed with applied fields, suggesting that LCO$_C$ layers are nonferromagnetic. These results are consistent with SQUID and NV magnetometry measurements. Microscopic magnetic imaging indicates that the lateral magnetic ground states in LCO hybrid homostructures strongly depend on epitaxial strain.

The correlation between strain and the electronic states of valence electrons in LCO hybrid homostructures was revealed by performing element-specific X-ray absorption spectroscopy (XAS) measurements at room temperature. Figure 3a shows the XAS results at O $K$-edges for LCO$_C$ and LCO$_T$ layers. The shadow region centered at ~530 eV represents the excitation of electrons from O 1$s$ to the Co 3$d$–O 2$p$ hybridization states. Both XAS curves agree well with the features of XAS at O $K$-edges for bulk LaCo$^{3+}$O$_3$ [26]. Meanwhile, the XAS results at Co $L$-edges showed strong peaks at ~780 eV ($L_3$) and ~795 eV ($L_2$) corresponding to the excitations of electrons from the $2p$ core levels to the $3d$ unoccupied states, consistent with the peak positions of $Co^{3+}$ ions [27]. These results indicate that the valence state of Co ions (+3) was maintained regardless of the film strain. Some layers did not show significant numbers of oxygen vacancies, which may influence the magnetization of LCO layers. Orbital occupancy in different regions of the LCO hybrid homostructure was further characterized by X-ray linear dichroism (XLD) using linearly polarized X-ray beams with variable incident angles, as shown in Figures 3b and 3c. When the incident angle is 90°, one anticipates that the absorption of X-rays ($I_{90°}$) arises entirely from in-plane orbitals ($d_{x^2-y^2}$). When the incident angle switches to 30° with respect to the surface plane, the XAS signals ($I_{30°}$) contain both $d_{x^2-y^2}$ and $d_{3z^2-r^2}$ orbital information. Comparing the difference in the peak energy between $I_{90°}$ and $I_{30°}$ (Figures 3d and 3e), we noticed that the $LCO_T$ layer under tensile strain had a lower peak energy of the $d_{x^2-y^2}$ orbital compared to that of the $d_{3z^2-r^2}$ orbital, while the $LCO_C$ layer showed the opposite effect. We calculated the nominal XLD value by subtracting $I_{30°}$ from $I_{90°}$ for both $LCO_T$ and $LCO_C$ layers (Figures 3f and 3g). The XLD value is negative for the tensile-strained $LCO_T$ layer, suggesting higher electron occupancy in $d_{x^2-y^2}$ orbitals, whereas the $LCO_C$ layer showed a positive XLD value, indicating that electrons preferentially occupy the $d_{3z^2-r^2}$ orbital instead. The difference in the electron occupancy determines the spin states of $Co^{3+}$ ions under different strain states. Consequently, the spin states of Co ions change from high spin states in $LCO_T$ (inset of Figure 3f) to low spin states in $LCO_C$ (inset of Figure 3g). The systematic XAS results provide solid evidence of the strain-mediated magnetic states of LCO hybrid homostructures, which agreed well with earlier magnetization characterizations.

To miniature ferromagnetic $LCO_T$ domains in the nonferromagnetic $LCO_C$ matrix, FS-STO nanodot arrays were fabricated top-down using a nanoporous anodic alumina mask and then thermally transferred onto LAO substrates [28]. Subsequently, LCO films were deposited on the modified substrates, as described in Methods. The schematic of the crossbar device structure with a vertical/horizontal write and readout geometry is shown in Figure 4a. The ferromagnetic signal can be read using spin transfer torque or thermally excited magnons through crossbars consisting of heavy metals or $5d$-element oxides with intrinsic large spin–orbital coupling effects [29,30]. In Figure 4b and Figure S6, we show $LCO_T$ nanodomains with a diameter of

~35 nm corresponding to an areal density of ~400 Gbit/in$^2$, which supersedes those of commercialized memory devices. A higher storage density could be achieved by further reducing the size of nanoislands. The experimentally achievable lower limit for the nanodomain size is unclear, which depends on growth conditions and nanofabrication techniques. However, the size of nanodomains would not greatly influence magnetic properties because the film thickness along the growth direction can be precisely controlled to maintain its long-range spin ordering. We analyzed strain distributions within LCO nanodomains along the in-plane and out-of-plane directions (Figures 4c and 4d). LCO$_T$ layers exhibited a larger in-plane lattice constant than LCO$_C$ layers, whereas the out-of-plane lattice constant of LCO$_T$ decreased slightly due to the tensile strain. Thus, strain modulation in LCO$_T$ nanoislands is valid at the nanoscale, implying that the large-area control of ordered nanoislands can promise various applications.

Finally, we further demonstrate that ferromagnetic nanodomains in cobaltite homostructures can be integrated into silicon-based CMOS technology. Using the same strategy, FS-STO membranes were transferred onto silicon (Figure 4e). The LCO$_T$ layers grown on FS-STO membranes were highly epitaxial and maintained a tensile strain of ~ 1.6%. The epitaxial strain of LCO$_T$ layers was slightly smaller than that of LCO$_T$ layers directly grown on STO substrates (Figure 4f). We observed a similar orbital polarization in LCO$_T$ layers independent of the target substrate (Figure S7), yielding identical strain effects on epitaxial LCO$_T$ layers. These LCO$_T$ layers showed a typical ferromagnetic character with a clear magnetic phase transition. $T_C$ increased by ~5 K compared to the $T_C$ of LCO$_T$ single layers. Moreover, the $M_S$ of LCO$_T$ layers on FS-STO membranes reached ~217 emu/cm$^3$ (~1.44 $\mu_B$/Co), which is ~50% larger than that of LCO$_T$ single layers. Enhanced $M_S$ in LCO$_T$ layers can be attributed to the reduced tensile strain. Previously, it was reported that the $M_S$ of LCO films grown on LSAT was larger than that of LCO films grown on STO [16, 31]. In both cases, the Co–O bond length ($d_{Co-O}$) was stretched, and the Co–O–Co bonding angle ($\beta_{Co-O-Co}$) remained at 180°. Slightly decreased tensile strain reduces $d_{Co-O}$. In this case, $(\Delta_{cf} - W/2) \propto (1/d_{Co-O}^5 - 1/d_{Co-O}^{3.5})$ continuously increases, leading to a strong overlap between the $t_{2g}$ and $e_g$ bands. Valence electrons depopulate the $t_{2g}$ band in favor of $d_{x^2-y^2}$ orbitals; thus, Co$^{3+}$ ions prefer high spin states and larger $M_S$ in LCO$_T$ layers on FS-STO membranes. Surprisingly, we observed a double hysteresis in *M–H* loops with a small coercive field of ~ 800 Oe. We believe the change in field-dependent magnetization is attributed to strain relaxation with increasing $LCO_T$ layer thickness. In addition, we measured the magnetic properties of the FS-LCO membrane and LCO amorphous layers grown directly on silicon (Figure S8). Both samples were nonferromagnetic, similar to their bulk form, revealing the influence of epitaxial strain on the stabilization of long-range spin ordering. The methodology described in the present work can be applied to arbitrary substrates, such as amorphous glass and α-$Al_2O_3$ (Figure S9). $LCO_T$ layers maintained their high crystallinity and orientation, similar to the buffered ultrathin membranes. This also allows the design of a functional grain boundary (GB) with dissimilar orientation and strain laterally. The atomically thin GB may serve as another type of information storage media with remarkably high density in the film plane, similar to long-term ferroelectric conductive domain wall memories. Further, this work enables the replacement of ultrathin dielectric STO membranes with other multifunctional (superconducting, ferroelectric, or photoelectric) oxide membranes. The stacking of these correlated oxide membranes with controllable twist angle, stacking order, and periodicity may provide a vigorous platform for both fundamental research and applied sciences [32].

**Discussion and conclusions**

The construction of ferromagnetic nanoislands in a nonmagnetic matrix based on cobaltites is demonstrated. The electronic and magnetic states of cobaltite homostructures were laterally modified by epitaxial strain, which was induced using membranes/substrates. A distinct magnetic contrast was obtained at the nanoscale, suggesting the capability to attain ultrahigh areal density for data storage. Furthermore, the generic method presented in this work is applicable to both silicon and flexible substrates. These results pave the way for the fabrication of nanoscale magnetic elements using distinct scenes for thin film epitaxy.

**Methods**

**Sample synthesis and X-ray based structural characterizations**

Pulsed laser deposition (PLD) technique was used to fabricate oxide thin films and heterostructures in the present work. We firstly fabricated the water-soluble $Sr_3Al_2O_6$ (SAO) layer with thickness ~ 30 nm on (001)-oriented $(LaAlO_3)_{0.3}$-$(Sr_2AlTaO_6)_{0.7}$ (LSAT) substrates (*Hefei Kejing Mater. Tech. Co. Ltd*). Subsequently, a 6-unit-cells-thick SrTiO$_3$ (STO) ultrathin layer was deposited on SAO layers. We kept the substrate's temperature was 650 °C and laser density was ~ 1.5 J/cm$^2$ for both layers. However, the oxygen partial pressure maintains 35 mTorr for SAO and increases to 100 mTorr for STO capping layers in order to achieve the best sample quality. After the sample growth, a thermal-release tape was firmly pressed on the as-grown samples and then immersed into de-ionized water at room temperature. The ultrathin freestanding (FS) STO membranes was adhered on the tape after the SAO fully dissolved. Then, the FS-STO membranes were carefully transferred on LaAlO$_3$ (LAO) substrates or other (glass, α-Al$_2$O$_3$, and silicon) supports using standard heating release process. Finally, LCO films with a thickness ~ 20 nm were fabricated on the modified substrates at the substrate's temperature of 750 °C, laser density of ~ 1 J/cm$^2$, and oxygen partially pressure of 200 mTorr. For the sample quality control purpose, we fabricated LCO single films on STO and LAO substrates for direct comparison. After the growth, the LCO samples were cooled down to room temperature under an annealing oxygen pressure of 100 Torr in order to largely remove the potential impacts from oxygen vacancies. The crystallographic analysis, including XRD θ-2θ scans and reciprocal space mapping (RSM), were carried out using synchrotron-based diffractometer and lab-based Panalytical X'Pert3 MRD diffractometer. The LCO film thickness (22 ± 1 nm) was obtained from fitting X-ray reflectivity curves using GenX software.

**Optical SHG polarimetry**

Optical SHG measurements were performed on a LCO hybrid structure. A focused 800 nm Ti: Sapphire femtosecond laser (Tsunami 3941-X1BB, Spectra-Physics) with a spot's diameter of 100 μm was incident at different regions of a LCO hybrid structure. The SHG signals were measured in the reflection geometry with incident angle of 45°. The second harmonic fields generated through the nonlinear optical process within LCO hybrid structures were decomposed into *p*-($I_{p-out}$) and *s*-($I_{s-out}$) polarized components by a beam splitter. The optical signals changes systematically with polarization direction (φ) of incident light. Then, the SHG polarimetry data were theoretically fitted with analytical models using standard point group symmetries.

**STEM imaging of hybrid structure**

Cross-sectional TEM specimens of LCO hybrid structures were prepared using $Ga^+$ ion milling after conventional mechanical thinning. The HAADF image were recorded in the scanning mode using JEM ARM 200CF microscopy at the Institute of Physics (IOP), Chinese Academy of Sciences (CAS). The sample is imaged from [100] zone axis of FS-STO membranes and LAO substrates. The atomic distances between A-cite elements were determined by fitting the intensity peaks using Gaussian function. The strain distribution within LCO hybrid structures were obtained using geometric pair analysis (GPA). All TEM data were analyzed using Gatan Digital Micrograph software.

**Magnetic characterizations**

The macroscopic magnetization of LCO hybrid structures were measured by a 7T-MPMS magnetometer. *M-T* curves were recorded during sample warm-up after field-cooled at 1 kOe. *M-H* loops were measured at 6 K by applying up to ± 7 T. The net magnetization of samples was obtained after subtracting diamagnetic signals from both FS-STO membranes and substrates. The evolution of magnetic domains was measured using a commercial cryogenic attocube MFM system. The hard magnetic coated point probes from Nano Sensor were used for scanning with a distance above the sample surface of ~ 100 nm. Magnetic fields up to ± 2 T were applied perpendicular to the sample surface using superconducting magnet. All MFM measurements were conducted at 6 K to ensure LCO hybrid stays in ferromagnetic state. The diamond nitrogen vacancy (NV)-based magnetometry measurements were performed at zero magnetic field using a home-built optical detected magnetic resonance (ODMR) system. The nanodiamonds (ND) with a diameter of ~ 100 nm were dispersed randomly on the surface of LCO hybrid structures. The ODMR signals were detected from $LCO_T$ and $LCO_C$ regions of an identical LCO hybrid sample separately by targeting an individual ND from corresponding regions using confocal microscopy. The measurements were taken by sweeping the microwave frequency through resonance and recording the optical excitation from an individual ND. The ODMR results were averaged over at least four measurements and double-checked at different NDs at the same region. The full ODMR spectrum exhibits a splatted two-peak-feature due to the Zeeman effect and can be fitted using Lorentz function. The ODMR intensity changes systematically with sweeping microwave frequency. The LCO hybrid sample was cooled down to 6 K using a closure He-recycling refrigerator. The measurements were performed progressively during warm-up process from 6 to 150 K. After each measurement, the sample was thermally stabilized for an hour to achieve the accurate temperature.

**XAS and XLD measurements**

Elemental specific XAS measurements were performed on LCO hybrid samples grown on LAO substrates at the beamline 4B9B of the Beijing Synchrotron Radiation Facility (BSRF). All spectra at both O $K$-edges and Co $L$-edges were collected at room temperature in total electron yield (TEY) mode. The LCO hybrid structures were properly grounded using copper tapes to obtain the best signal-to-noise ratio. The typical photocurrents measured using TEY mode were in the order of pA. XAS measurements were performed alternately at $LCO_C$ and $LCO_T$ regions of LCO hybrid structures. The incident angles of linearly polarized X-ray beam vary from 90° to 30° with respect to the surface plane. When the incident angle sets to 90°, XAS signals reflects the $d_{x^2-y^2}$ orbital occupancy directly, whereas the incident angle changes to 30°, XAS contains both $d_{x^2-y^2}$ and $d_{3z^2-r^2}$ orbital information. To quantify the orbital occupancy in Co $e_g$ bands, we calculated the nominal XLD by $I_{90°} - I_{30°}$. All XAS data were normalized to the values at the pre- and post-edges for direct comparison.

**Acknowledgements**

We thank Tengyu Guo at SLAB for the extensive assistance in magnetization measurements. **Funding**: This work was supported by the National Key Basic Research Program of China (Grant Nos. 2020YFA0309100 and 2019YFA0308500), the National Natural Science Foundation of China (Grant Nos. 11974390, U22A20263, 52250308, 11721404, 12174364, 11874412, 52072400, 52025025, and 12174347), the Guangdong-Hong Kong-Macao Joint Laboratory for Neutron Scattering Science and Technology, and the Strategic Priority Research Program (B) of the Chinese Academy of Sciences (Grant No. XDB33030200). Synchrotron-based XRD measurements were performed at the beamline 1W1A and XAS and XLD experiments were performed at the beamline 4B9B of the Beijing Synchrotron Radiation Facility (BSRF) via user proposals. **Author contributions:** The freestanding STO layers and LCO hybrid samples were grown and processed by S.R.C.; the transfer of freestanding oxide membranes was conducted by D.K.R.; Temperature dependent NV magnetometry was performed by Y.X. and Y.X.S. under guidance of G.Q.L.; MFM images were recorded by M.M.C. under guidance of J.X.Z.; SHG measurements were carried out by S.X. under guidance of K.J.J.; XAS and XLD measurements were performed by S.R.C., H.T.H., T.C., Q.J. and J.O.W.; TEM lamellas were fabricated with FIB milling and TEM experiments were performed byX.Y.L., Q.H.Z. and L.G.; S.R.C., D.K.R., T.C., Q.J., and H.T.H. worked on the structural and magnetic measurements. C.W. and H.G. participated the discussions and K.J.J. provided important suggestions during the manuscript preparation. E.J.G. initiated the research and supervised the work. S.R.C. and E.J.G. wrote the manuscript with inputs from all authors. **Competing interests:** The authors declare that they have no competing financial interests. **Data and materials availability:** The data that support the findings of this study are available on the proper request from the first author (S.R.C.) and the corresponding authors (E.J.G.). **Additional information**: Supplementary information is available in the online version of the paper. Reprints and permissions information is available online.

## References and notes

[1] S.N. Piramanayagam and Tow C. Chong, *Developments in Data Storage*, IEEE-Wiley Press (2012).

[2] A. von Reppert, L. Willig, J.-E. Pudell, S. P. Zeuschner, G. Sellge, F. Ganss, O. Hellwig, J. A. Arregi, V. Uhlíř, A. Crut, M. Bargheer. Spin stress contribution to the lattice dynamics of FePt. Science Advances, 6, eaba1142 (2020).

[3] A. Singh, et al. Phys. Rev. Lett. 105, 067206 (2010).

[4] E. Berganza et al., Half-hedgehog spin textures in sub-100 nm soft magnetic nanodots. Nanoscale 12, 18646-18653 (2020).

[5] David E. J. Waddington, Thomas Boele, Richard Maschmeyer, Zdenka Kuncic, Matthew S. Rosen, High-sensitivity in vivo contrast for ultra-low field magnetic resonance imaging using superparamagnetic iron oxide nanoparticles, Sci. Adv. 6, eabb0998 (2020).

[6] M. Mallary; et al., One terabit per square inch perpendicular recording conceptual design, IEEE Transactions on Magnetics. 38 (4): 1719–1724 (2002).

[7] Christoph Vogler et al., Heat-assisted magnetic recording of bit-patterned media beyond 10 Tb/in$^2$, Appl. Phys. Lett. 108, 102406 (2016).

[8] J. Mannhart, and D. G. Schlom, Oxide interfaces-An opportunity for electronics, Science

327, 1607 (2010).

[9] D. Kan et al., Tuning magnetic anisotropy by interfacially engineering the oxygen coordination environment in a transition metal oxide, Nat. Mater. 15, 432 (2016).

[10] Z. Liao et al., Controlled lateral anisotropy in correlated manganite heterostructures by interface-engineered oxygen octahedral coupling, Nat. Mater. 15, 425 (2016).

[11] S. Lin, et al., Switching magnetic anisotropy of $SrRuO_3$ by capping-layer-induced octahedral distortion, Phys. Rev. Appl. 13, 034033 (2020).

[12] J. Zhang et al., Symmetry mismatch-driven perpendicular magnetic anisotropy for perovskite/brownmillerite heterostructures, Nat. Commun. 9, 1923 (2018).

[13] S. Thomas et al., Localized control of Curie temperature in perovskite oxide film by capping layer induced octahedral distortion, Phys. Rev. Lett. 119, 117203 (2017).

[14] D. Fuchs et al., Ferromagnetic order in epitaxial strained $LaCoO_3$ thin films, Phys. Rev. B 75, 144402 (2007).

[15] V. V. Mehra et al., Long-range ferromagnetic order in $LaCoO_{3-\delta}$ epitaxial films due to the interplay of epitaxial strain and oxygen vacancy ordering, Phys. Rev. B 91, 144418 (2015).

[16] W. S. Choi et al., Strain-induced spin states in atomically ordered cobaltites, Nano Lett. 12, 4966-4970 (2012).

[17] R. J. Zeches, et al., A strain-driven morphotropic phase boundary in $BiFeO_3$, Science 326, 977–980 (2009).

[18] A. Herklotz et al., Wide-range strain tunability provided by epitaxial $LaAl_{1-x}Sc_xO_3$ template films, New J. Phys. 12, 113053 (2010).

[19] P. -C. Wu, C. -C. Wei, Q. Zhong, S. -Z. Ho, Y. -D. Liou, Y. -C. Liu, C. -C. Chiu, W. -Y. Tzeng, K. -E. Chang, Y. -W. Chang, J. Zheng, C. -F, Chang, C. -M. Tu, T. -M. Chen, C. -W. Luo, R. Huang, C. -G. Duan, Y. -C. Chen, C. -Y. Kuo, J. -C. Yang, Twisted oxide lateral homostructures with conjunction tunability, Nat. Commun. 13, 2565 (2022).

[20] S. Chen et al., Braiding Lateral Morphotropic grain boundaries in homogenetic oxides, Adv. Mater. 35, 2206961 (2023).

[21] D. Lu, D. J. Baek, S. S. Hong, L. F. Kourkoutis, Y. Hikita, H. Y. Hwang, Nat. Mater. 15, 1255 (2016).

[22] J.-H. Kwon, W. S. Choi, Y.-K. Kwon, R. Jung, J.-M. Zuo, H. N. Lee, and J. Kim, Nanoscale

spin-state ordering in LaCoO$_3$ epitaxial thin films, Chem. Mater., 26, 2496 (2014).

[23] E. J. Guo et al., Exploiting Symmetry Mismatch to Control Magnetism in a Ferroelastic Heterostructure, Phys. Rev. Lett. 122, 187202 (2019).

[24] E. J. Guo et al., Nanoscale ferroelastic twins formed in strained LaCoO$_3$ films, Sci. Adv. 5, eaav5050 (2019).

[25] V. M. Acosta et al., Temperature Dependence of the Nitrogen-Vacancy Magnetic Resonance in Diamond, Phys. Rev. Lett. 104, 070801 (2010).

[26] Z. Hu, H. Wu, M. W. Haverkort, H. H. Hsieh, H. J. Lin, T. Lorenz, J. Baier, A. Reichl, I. Bonn, C. Felser, A. Tanaka, C. T. Chen, and L. H. Tjeng, Different Look at the Spin State of Co$^{3+}$ Ions in a CoO$_5$ Pyramidal Coordination, Phys. Rev. Lett. 92, 207402 (2004).

[27] M. W. Haverkort, Z. Hu, J. C. Cezar, T. Burnus, H. Hartmann, M. Reuther, C. Zobel, T. Lorenz, A. Tanaka, N. B. Brookes, H. H. Hsieh, H.-J. Lin, C. T. Chen, and L. H. Tjeng, Spin state transition in LaCoO$_3$ studied Using Soft x-ray Absorption Spectroscopy and Magnetic Circular Dichroism, Phys. Rev. Lett. 97, 176405 (2006).

[28] Zhongwen Li, Yujia Wang, Guo Tian, Peilian Li, Lina Zhao, Fengyuan Zhang, Junxiang Yao, Hua Fan, Xiao Song, Deyang Chen, Zhen Fan, Minghui Qin, Min Zeng, Zhang Zhang, Xubing Lu, Shejun Hu, Chihou Lei, Qingfeng Zhu, Jiangyu Li, Xingsen Gao, Jun-Ming Liu, High-density array of ferroelectric nanodots with robust and reversibly switchable topological domain states, Science Adv. 3, e1700919 (2017).

[29] Xinde Tao, Qi Liu, Bingfeng Miao, Rui Yu, Zheng Feng, Liang Sun, Biao You, Jun Du, Kai Chen, Shufeng Zhang, Luo Zhang, Zhe Yuan, Di Wu, Haifeng Ding, Self-consistent determination of spin Hall angle and spin diffusion length in Pt and Pd: The role of the interface spin loss, Sci. Adv. 4, eaat1670 (2018).

[30] Jobu Matsuno, Naoki Ogawa, Kenji Yasuda, Fumitaka Kagawa, Wataru Koshibae, Naoto Nagaosa, Yoshinori Tokura, Masashi Kawasaki, Interface-driven topological Hall effect in SrRuO$_3$-SrIrO$_3$ bilayer, Sci. Adv. 2, e1600304 (2016).

[31] E. J. Guo et al., Switchable orbital polarization and magnetization in strained LaCoO$_3$ films, Phys. Rev. Mater. 3, 014407 (2019).

[32] Y. Cao, V. Fatemi, A. Demir, S. Fang, S. L. Tomarken, J. Y. Luo, J. D. Sanchez- Yamagishi, K. Watanabe, T. Taniguchi, E. Kaxiras, R. C. Ashoori, P. Jarillo-Herrero, Nature 556, 80 (2018).

**Figures and figure captions**

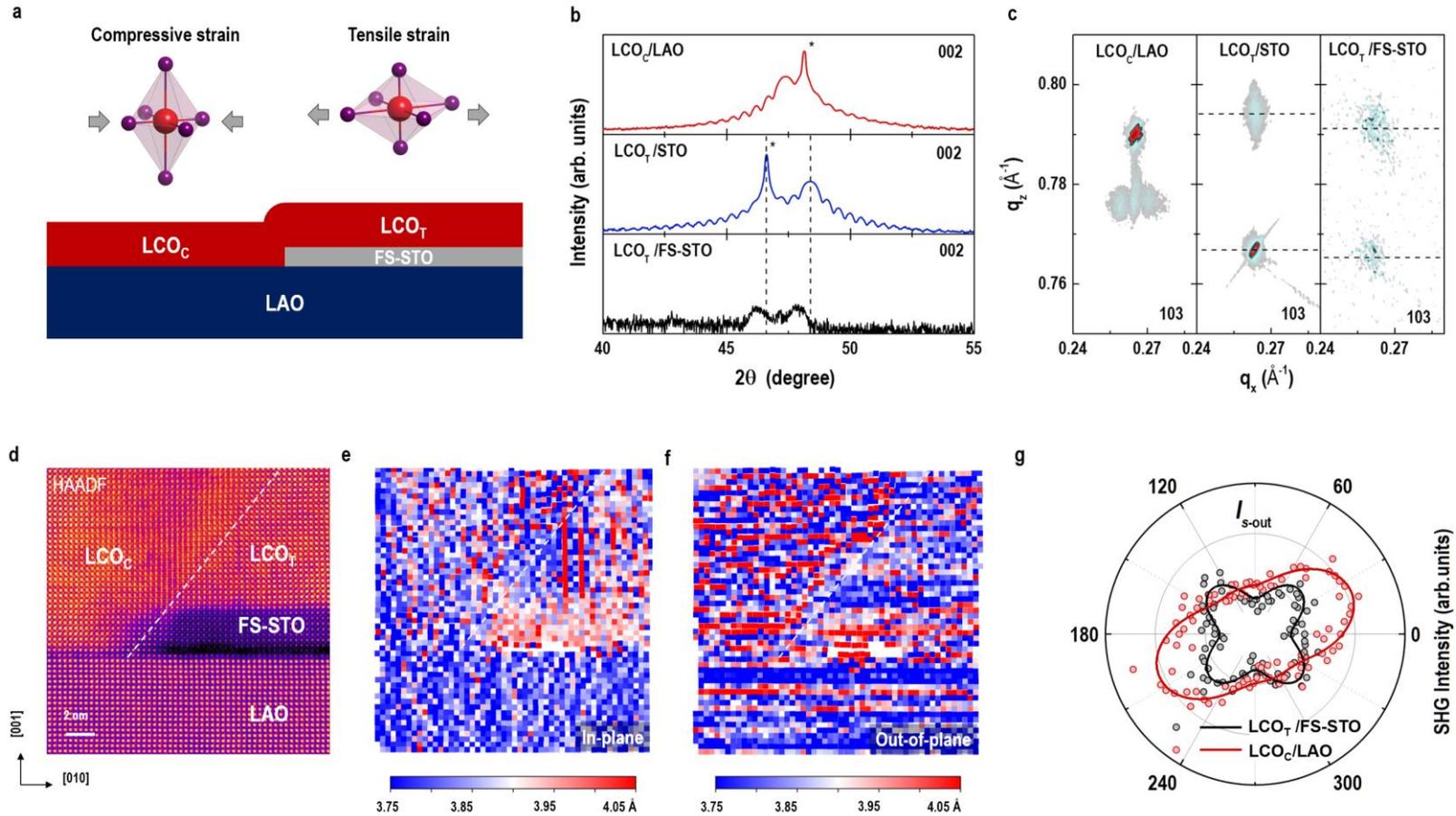

**Figure 1. Structural characterizations of an LCO hybrid structure**. (a) Schematic of an LCO hybrid structure, which composes a compressively strained LCO (LCO$_C$) region (on LAO) and a tensile-strained LCO (LCO$_T$) region (on FS-STO). (b) XRD θ–2θ scans of LCO single layers grown on LAO and STO substrates, as well as FS-STO membranes. (c) RSM of these LCO single layers around substrates/membranes' (103) reflection. (d) Cross-sectional HAADF-STEM image at a respective grain boundary in an LCO hybrid structure. (e) and (f) Calculated in-plane and out-of-plane atomic distances, respectively. (g) SHG polarimetry of $I_{s\text{-out}}$ signals from LCO$_C$ and LCO$_T$ regions in an LCO hybrid structure.

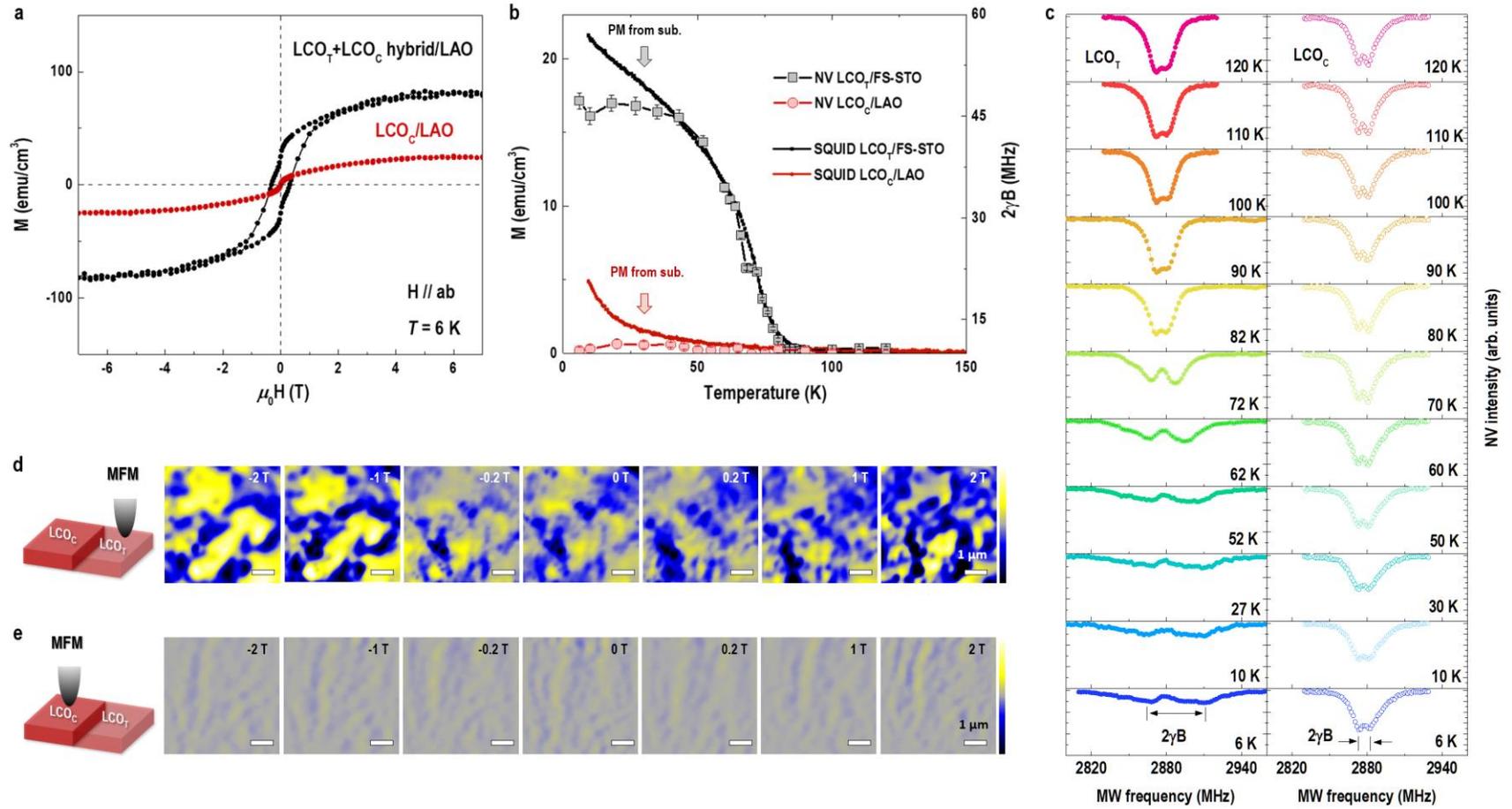

**Figure 2. Magnetically distinct regions in an LCO hybrid structure**. (a) *M–H* hysteresis loops and (b) *M–T* curves of an $LCO_C$ single layer and an LCO hybrid on LAO substrates. *M–H* loops were recorded at 6 K. *M–T* curves were measured during sample warm-up under an in-plane field of 1 kOe. (c) Zero-field ODMR spectra of NV centers dispersed on the LCO hybrid structure. ODMR spectra were collected from $LCO_T$ (left) and $LCO_C$ (right) regions, respectively. $2\gamma B$ were subtracted and plotted in (b) as open symbols. (d) and (e) MFM images taken from $LCO_T$ and $LCO_C$ regions at 6 K, respectively, under different external magnetic fields from –2 to 2 T. The scale bar is 1 μm.

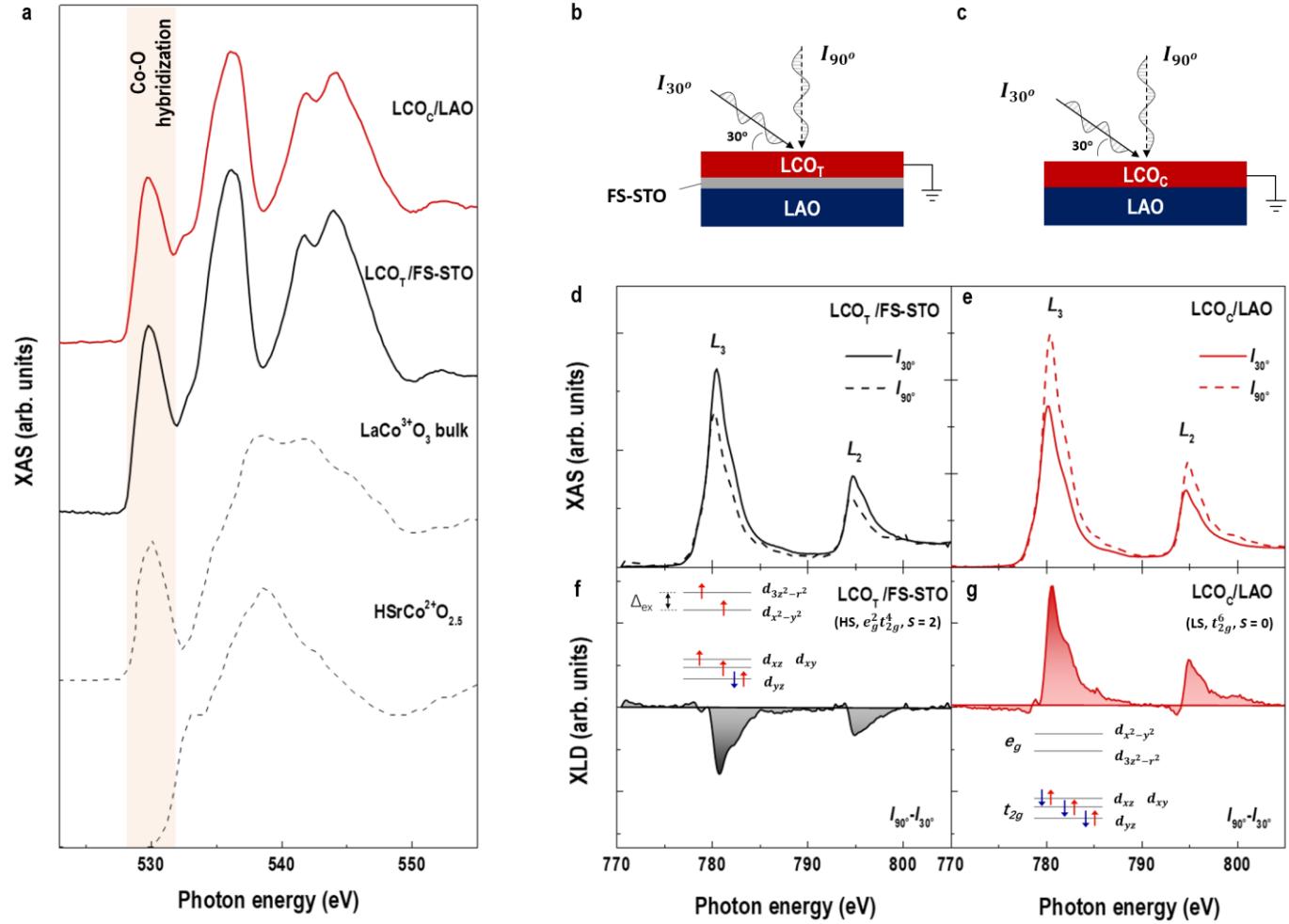

**Figure 3. Laterally separated spin states in an LCO hybrid structure**. (a) XAS at O $K$-edges collected from $LCO_C$ and $LCO_T$ regions. The dashed lines show the reference data from bulk $LaCo^{3+}O_3$ and $HSrCo^{2+}O_{2.5}$ films. (b) and (c) Schematics of measured geometries in $LCO_C$ and $LCO_T$ regions, respectively. XAS were collected using the TEY mode at room temperature. (d) and (e) XAS at Co $L$-edges from $LCO_C$ and $LCO_T$ regions, respectively. The calculated nominal XLD (= $I_{90°} - I_{30°}$) of $LCO_C$ and $LCO_T$ regions are shown in (f) and (g), respectively.

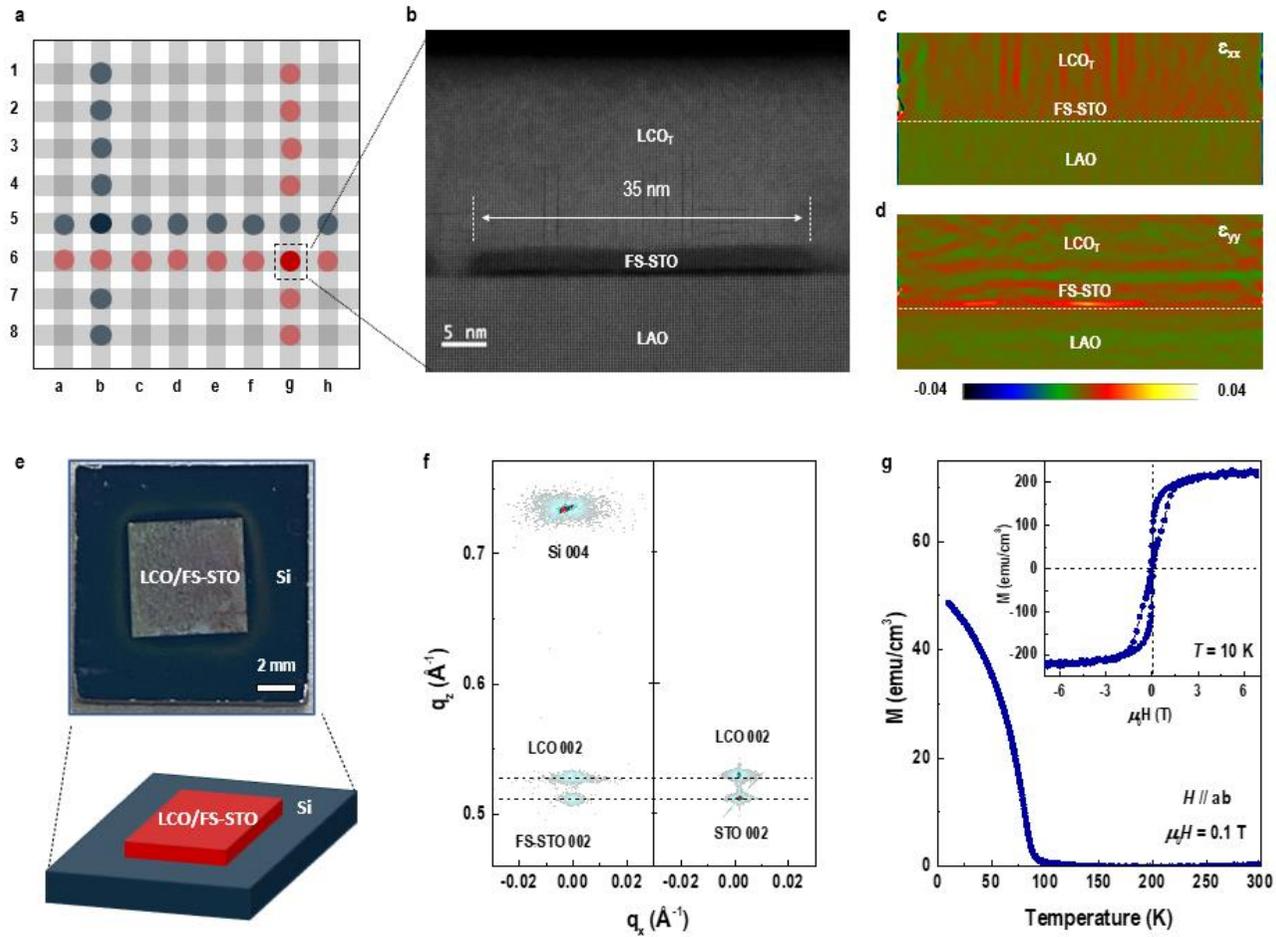

**Figure 4. Ferromagnetic nanoislands in a nonmagnetic matrix**. (a) Schematic of programmatic addressable magnetic nanoislands. (b) Cross-sectional HAADF-STEM image of a single $LCO_T$ nanodomain in the $LCO_C$ matrix. (c) and (d) In-plane strain ($\varepsilon_{xx}$) and out-of-plane strain ($\varepsilon_{yy}$) distributions at the interface, respectively. (e) Schematic and optical photograph of an $LCO_T$/FS-STO membrane transferred on silicon. (f) Direct comparison of RSMs between an $LCO_T$/FS-STO membrane on silicon and an $LCO_T$ film on STO around substrates' (00L) reflections. (g) $M$–$T$ curve of an $LCO_T$/FS-STO membrane on silicon under an in-plane field of 1 kOe. The inset shows its $M$–$H$ loops at 10 K.

# Supplementary Materials for

## Miniature Magnetic Nanoislands in a Morphotropic Cobaltite Matrix

Shengru Chen,† Dongke Rong,† Yue Xu, Miming Cai,, Qinghua Zhang, Shuai Xu, Yan-Xing Shang, Haitao Hong, Ting Cui, Qiao Jin, Jia-Ou Wang, Haizhong Guo, Lin Gu, Qiang Zheng, Can Wang, Jinxing Zhang, Gang-Qin Liu, Kui-juan Jin,* and Er-Jia Guo*

Correspondence to: E-mail: kjjin@iphy.ac.cn and ejguo@iphy.ac.cn
†These authors contribute equally to the manuscript.

**This PDF file includes:**

**Figs. S1 to S9.**

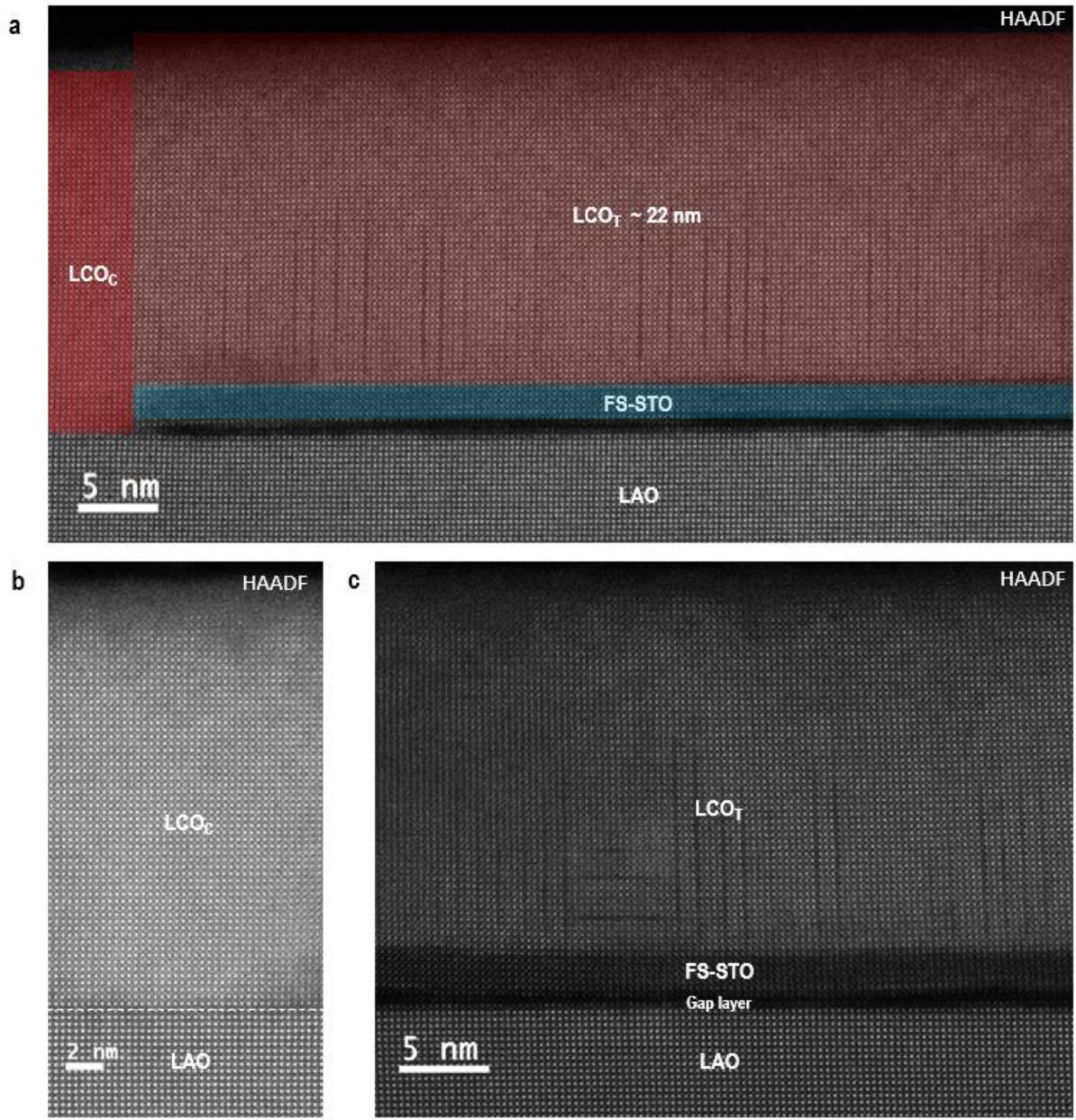

**Figure S1. Microscopic structural characterization of a LCO hybrid structure**. High-magnified cross-sectional HAADF-STEM images of (a) a LCO hybrid structure at a representative grain boundary region, (b) a $LCO_C$ layer grown directly on LAO substrates, and (c) a $LCO_T$ layer epitaxially grown on FS-STO membranes transferred on LAO substrates. The thickness of $LCO_C$ and $LCO_T$ layers is approximately 22 nm. The FS-STO membranes with a thickness of ~ 6 unit cells are continuous and flat. It covers partial surface of LAO substrates, leaving a space gap layer with thickness less than 1 nm. Apparently, the $LCO_C$ layers are absent from dark strips, where the $LCO_T$ layers contain vertical aligned dark strips. This behavior demonstrates that the LCO layers exhibit distinct strain states laterally depending on their substrates/membranes underneath. Please note that the dark strips in $LCO_T$ layers only persist approximately one half of entire layers. This case is different from previous work that the dark stripes occupy the entire LCO single films grown on STO substrates. We attribute this behavior to the partial relaxation of epitaxial strain provided by FS-STO membranes.

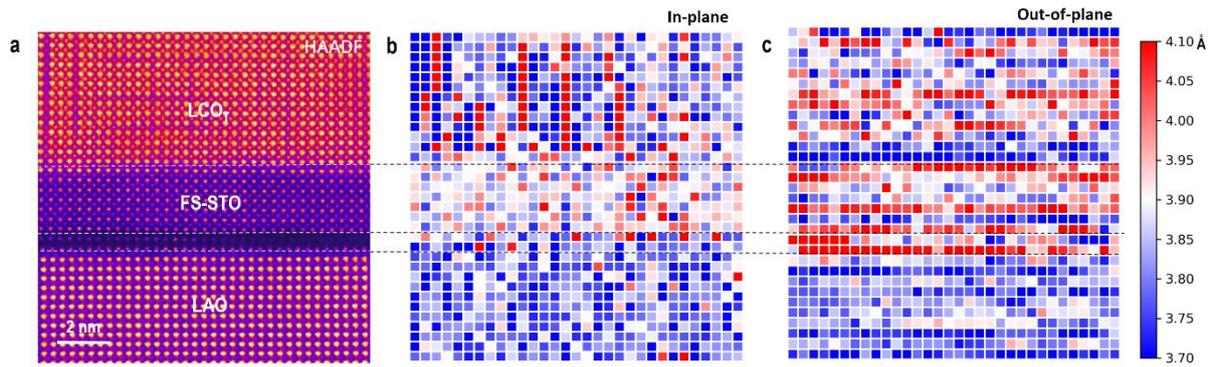

**Figure S2. Strain analysis of LCO$_T$ layer grown on FS-STO membranes**. (a) High-magnified HAADF-STEM image at LCO$_T$/FS-STO/LAO interface region. The black area is the interfacial gap layer between FS-STO membranes and LAO substrates. (b) and (c) Atomic distance between nearby A-cite elements along the in-plane and out-of-plane direction, respectively. We found that the out-of-plane lattice constant of FS-STO is larger than its in-plane lattice constant. Obviously, this result can only be attributed to the as-grown LCO$_T$ layers reversely apply a compressive-strain to the ultrathin FS-STO membranes. Meanwhile, the in-plane lattice constant of LCO$_T$ layers is slightly smaller than the out-of-plane ones.

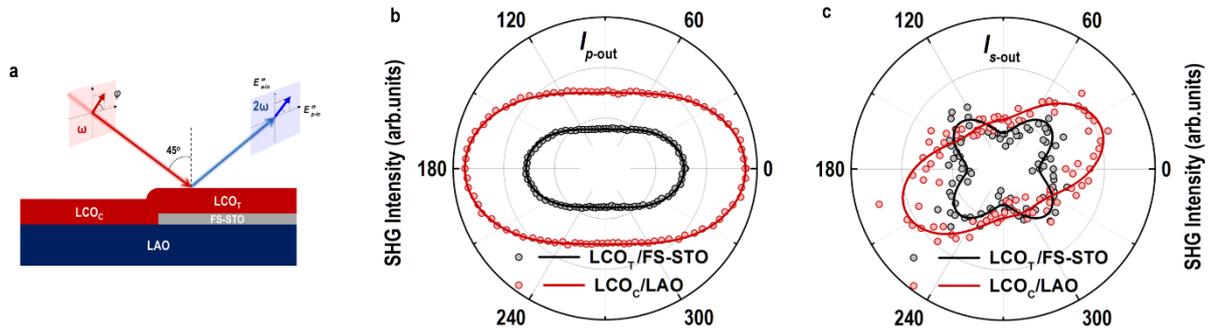

**Figure S3. SHG polarimetry of a LCO hybrid structure**. (a) Schematic of SHG measurements performed in reflection mode. The incident angle of 800 nm fs-laser is set to 45º with respect to the sample's surface normal. The measurements were repeated in both LCO$_C$ and LCO$_T$ regions at room temperature. (b) and (c) $I_{p\text{-out}}$ and $I_{s\text{-out}}$ components of SHG signal were plotted as function of polarization angle ($\varphi$), respectively. Red and black open symbols represent the experimental data obtained from LCO$_C$ and LCO$_T$ regions, respectively. Solid lines are the best fit to the SHG data and plotted for guide. We find that the SHG signal of LCO$_T$ layers can fit the point group symmetry *P4mm*, where that of LCO$_C$ fit the point group symmetry *m* better.

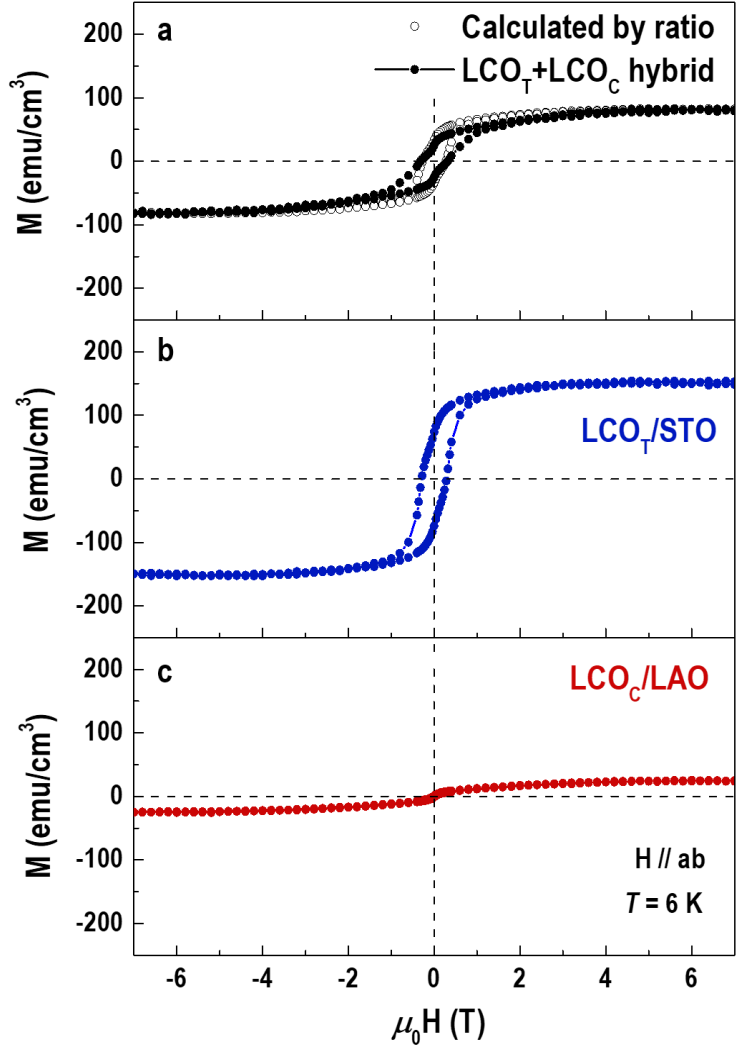

**Figure S4. Magnetic hysteresis loops** of (a) a LCO hybrid structure, (b) a $LCO_T$ single film, and (c) a $LCO_C$ single film. *M-H* curves were recorded at 6 K under in-plane magnetic fields. The $LCO_T$ single film exhibits a typical ferromagnetic behavior, whereas the $LCO_C$ single film does not show an apparent open hysteresis loop and saturation moment is small. These results agree with previous reports that the LCO films show ferromagnetism under tensile strain, while it exhibits suppressed long-range spin ordering under compressive strain. We further compare the averaged magnetization of a LCO hybrid structure with calculated magnetization weighted by area ratios of $LCO_T$ (~ 40%) and $LCO_C$ (~ 60%) layers. Both results are in reasonably good agreement in both coercive fields and saturation magnetization.

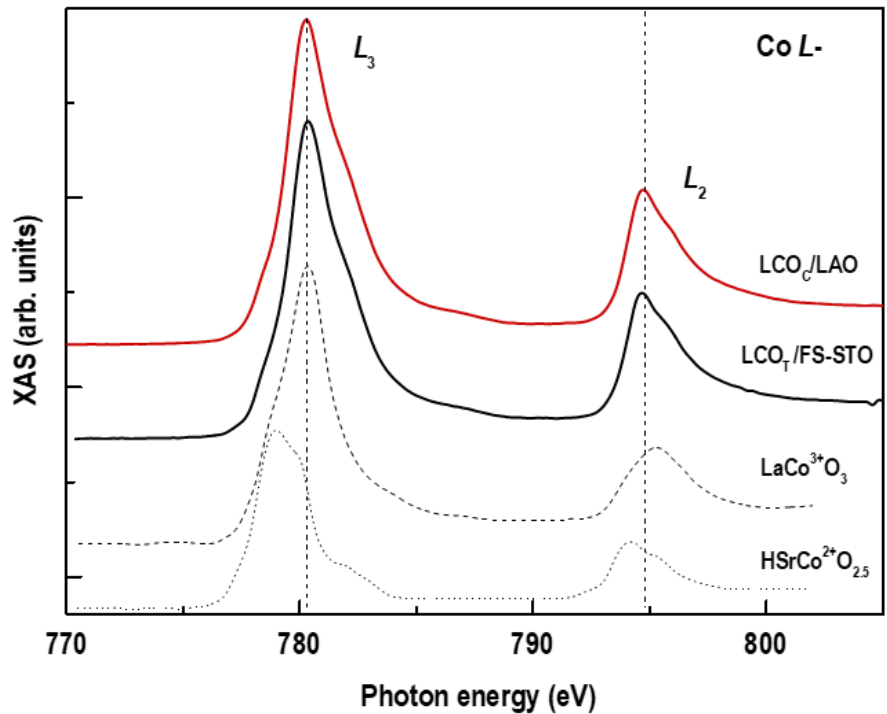

**Figure S5. XAS at Co *L*-edges for LCO$_C$/LAO and LCO$_T$/FS-STO heterostructures.** The spectra from different samples are shifted for clarification. The reference XAS Co *L*-edge data obtained from a LCO bulk and a HSrCoO$_{2.5}$ thin film were plotted as dashed lines. For LCO$_C$ and LCO$_T$ layers show strong XAS peaks at ~ 780.5 and 795 eV, in consistent with peak positions of the Co $L_3$- and $L_2$-edges for Co$^{3+}$ ions, respectively. The valence state of Co ions is independent from strain states of LCO films, agreeing with our previous work. All spectra were collected at room temperature in TEY mode.

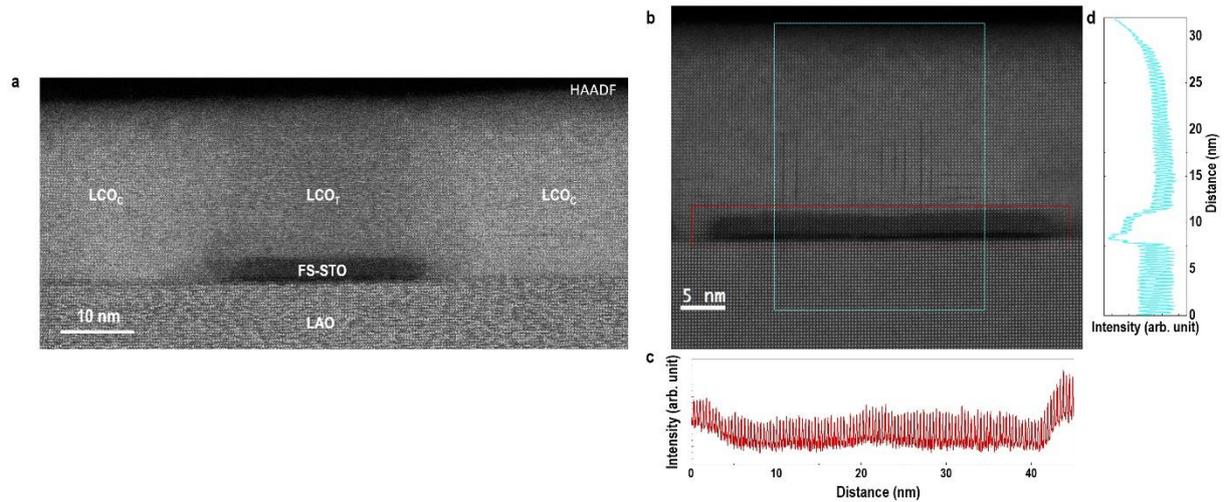

**Figure S6. Structural analysis of a representative LCO nano-island**. (a) Low-magnified HAADF-STEM image of an artificial LCO nano-island grown on LAO substrates. Except for FS-STO membranes, the sample show bright contrast in the STEM image due to the stronger electron scattering from La atoms. (b) Zoom-in HAADF-STEM image of an artificial LCO nano-island. Both vertical and horizontal dark stripes with a period of 3-unit-cells are observed at the $LCO_T$ regions. (c) and (d) Intensity profiles along the in-plane and out-of-plane direction, respectively. From in-plane intensity profile, we could determine the diameter of $LCO_T$ nano-island is $(35 \pm 2)$ nm. The thickness of FS-STO membranes is ~ 3 nm (approximately 6-7 unit cells). The gap between FS-STO membranes and LAO substrates is less than 1 nm. The structural characterizations confirm the miniature of ferromagnetic $LCO_T$ nano domains surrounded by non-ferromagnetic $LCO_C$ domains.

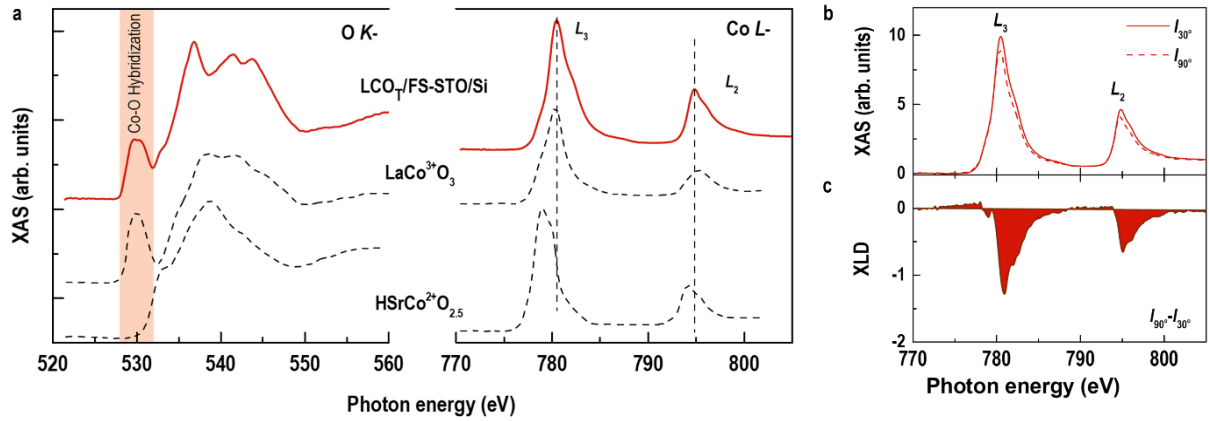

**Figure S7. Electronic state of LCO$_T$/FS-STO membranes transferred on silicon**. (a) XAS at O $K$- and (b) Co $L$-edges for LCO$_T$/FS-STO/Si sample. Reference data for Co$^{3+}$ and Co$^{2+}$ ions and their hybridization with oxygen ions were shown as dashed lines. Epitaxial growth of LCO$_T$ on FS-STO membranes does not change the valence state (+3) of Co ions. (b) XAS at Co $L$-edges when X-ray beam is parallel to the surface normal ($I_{90°}$, solid line) or has an incident angle of 30° with respect tot the surface normal ($I_{30°}$, dashed line). XAS represents directly the orbital occupancy in $e_g$ bands. XAS ($I_{90°}$) reflects the occupancy of Co $d_{3z^2-r^2}$ orbital, while XAS($I_{30°}$) contains both orbital information of in-plane and out-of-plane orientations. The peak energy for XAS($I_{30°}$) is lower than that of XAS ($I_{90°}$), suggesting the in-plane orbitals have lower energy than the out-of-plane ones. (c) Calculated XLD ($=I_{90°}-I_{30°}$) for LCO$_T$/FS-STO/Si sample. The negative value of XLD demonstrates the electrons are in favor of occupying the $d_{x^2-y^2}$ orbitals rather than the $d_{3z^2-r^2}$ orbitals due to the lower energy of $d_{x^2-y^2}$ orbitals in tensile-strained LCO$_T$ layers.

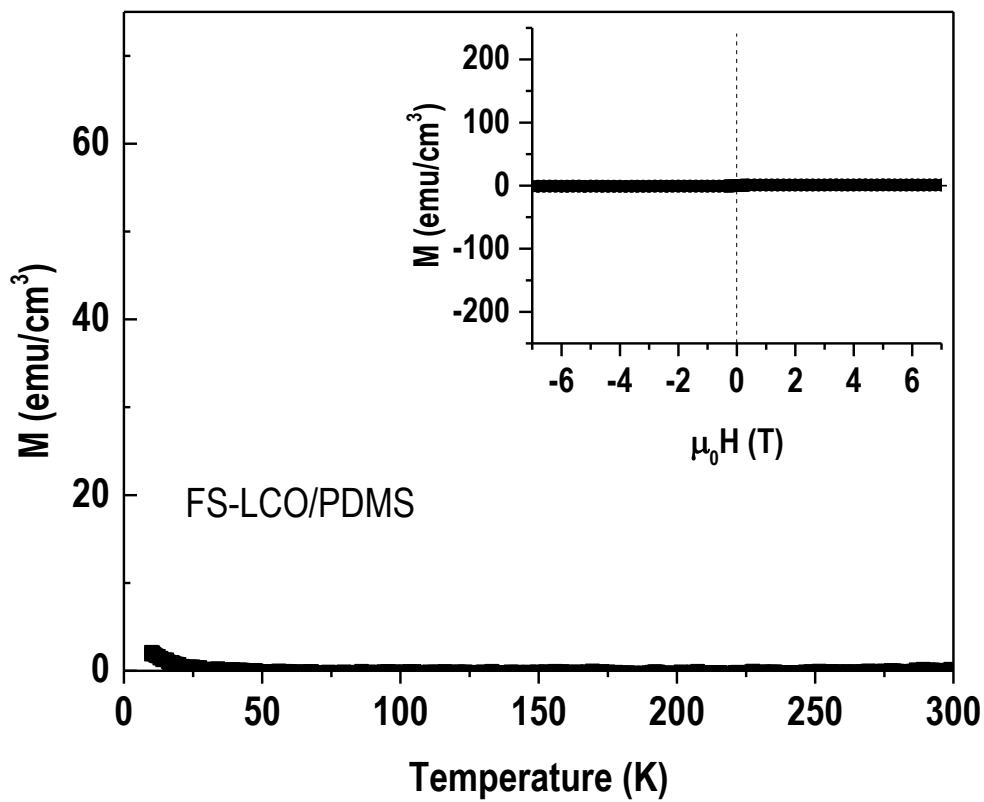

**Figure S8. Magnetic properties of FS-LCO membranes on PDMS**. *M-T* curve of FS-LCO membranes demonstrates no magnetic phase transition as increasing temperature. Inset shows the *M-H* curve of FS-LCO membranes. No magnetic hysteresis loop is observed. The magnetization measurements indicate FS-LCO membranes are not ferromagnetic, similar to its bulk form.

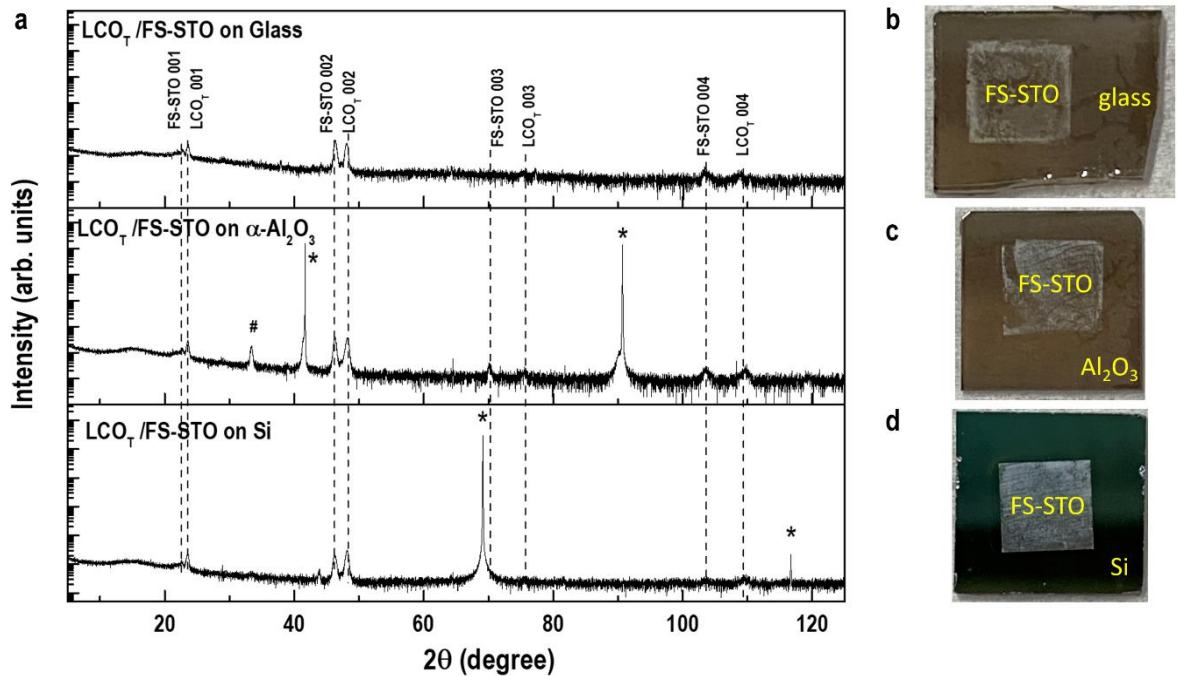

**Figure S9. Epitaxy of LCO$_T$ on FS-STO membranes transferred to arbitrary substrates**. (a) XRD θ-2θ scans of LCO$_T$ grown on 10nm-thick FS-STO modified glass, α-Al$_2$O$_3$, and silicon substrates. Dashed lines mark the 00$l$ peak positions of LCO$_T$ layers and FS-STO membranes. "*" denotes the substrate's peaks. "#" in the middle curve identifies an undefined impurity peak. All results reinforce the epitaxial growth of LCO$_T$ layers on FS-STO membranes although the rest of LCO layers grown directly on arbitrary substrates are amorphous due to the large lattice mismatch or difference crystalline symmetries.